\documentclass[conference]{IEEEtran}
\ifCLASSINFOpdf
  \usepackage[pdftex]{graphicx}
\else
\fi
\usepackage{amsmath}
\usepackage{ctable}

\usepackage{url}

\begin{document}
%
\title{Densely Connected CNNs for Bird Audio Detection}

\author{\IEEEauthorblockN{Thomas Pellegrini}
\IEEEauthorblockA{IRIT, Universit\'e de Toulouse, Toulouse, France\\
Email: thomas.pellegrini@irit.fr}}

\maketitle

\begin{abstract}
Detecting bird sounds in audio recordings automatically, if accurate enough, is expected to be of great help to the research community working in bio- and ecoacoustics, interested in monitoring biodiversity based on audio field recordings. To estimate how accurate the state-of-the-art machine learning approaches are, the Bird Audio Detection challenge involving large audio datasets was recently organized. In this paper, experiments using several types of convolutional neural networks (i.e. standard CNNs, residual nets and densely connected nets) are reported in the framework of this challenge. DenseNets were the preferred solution since they were the best performing and most compact models, leading to a 88.22\% area under the receiver operator curve score on the test set of the challenge (ranked $3^{\textrm{rd}}/30$)\footnote{Challenge solution source code available at \url{https://github.com/topel/bird_audio_detection_challenge}}. Performance gains were obtained thank to data augmentation through time and frequency shifting, model parameter averaging during training and ensemble methods using the geometric mean. On the contrary, the attempts to enlarge the training dataset with samples of the test set with automatic predictions used as pseudo-groundtruth labels consistently degraded performance.
\end{abstract}


%
\IEEEpeerreviewmaketitle

\section{Introduction}

Automatic detection of animal vocalizations, such as singing birds, besides being a scientific challenge in itself, can be helpful for monitoring biodiversity. Researchers involved in bioacoustics and/or the recent field named ecoacoustics~\cite{sueur_farina_2015} gather ever-growing quantities of \textit{in-situ} recordings that need to be manually analyzed. Automatic tools that label the recordings accurately are very much in demand to ease the time-consuming task of listening to hours of them~\cite{guyot2016sinusoidal}. 

Bird sounds are a topic of intensive research given their richness and variety, but also due to the fact that birds are more easily detectable through the audio modality rather than vision. Bird species' identification have been the target of several international evaluation campaigns such as LifeCLEF (BirdCLEF), a yearly contest including bird species identification in \textit{in-situ} audio recordings\footnote{\url{http://www.imageclef.org/}}. A variety of machine learning techniques have been explored for this task. For a complete survey, the reader may refer to \cite{stowell2016bird}, which also describes the Bird Audio Detection (BAD) challenge in which the work described in the present paper took place. The winning solutions of last year BirdCLEF edition were based on deep convolutional neural networks (CNNs)~\cite{sprengel2016audio}. Indeed, the success of \textit{deep learning} (DL) and deep neural networks (DNN) in many domains involving classification tasks, offers new and appealing perspectives.

In the context of the BAD challenge, concerned with the detection of bird sound in short duration recordings, the solution reported in this paper is based on CNNs, more specifically on densely connected CNNs, also called denseNets~\cite{huang2016densely}. The paper is organized as follows. First, brief descriptions of the BAD challenge task and dataset partitions are given, followed by sections on features, models and experiments carried out on the challenge data. Finally, Section \ref{sec:saliency} is an attempt to broaden the technical scope of the paper by describing the generation of "saliency maps", used as a visualization tool and a way to re-synthesize the original audio samples with strengthened time-frequency blobs salient for the networks.

\begin{table}
\begin{center}
\caption{Dataset partitions given in number of files}
\label{tab:data}
\begin{tabular}{lccc}
\toprule
 & Train & Valid & Test \\
\cmidrule(lr){2-4}
Freefield1010 (FF) & \hphantom{1}6,152 & 384 & 1,154\\
Warblr (W) & \hphantom{1}6,800 & 500 & \hphantom{1,}700 \\ 
\midrule
Freefield1010 + Warblr (FF\_W\_1) & 12,952 & 884 & 1,854\\ 
Freefield1010 + Warblr (FF\_W\_2) & 14,806 & 884 & \\ 
\midrule
Official test set &  &  & 8,620 \\
\bottomrule
\end{tabular}
\end{center}
\end{table}

\begin{figure*}
\centering
\includegraphics[scale=0.59]{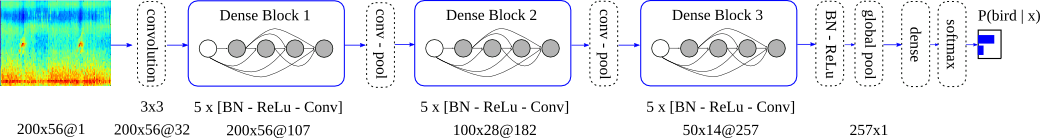}
\caption{DenseNet architecture.}
\label{fig:densenet}
\end{figure*}

\section{Challenge overview}

The BAD challenge's task was to design automatic systems that, given a short audio recording, returns a binary decision for the presence/absence of bird sound~\cite{stowell2016bird}. The participants were provided with two large development corpus and had to provide predictions of the presence of bird sound on the 8,620 files comprising an official held-out test set. Each team could submit up to 20 predictions. The official performance metric was the Area Under the Curve (AUC) score calculated on the receiver operator curve (ROC). For each submission, a preview estimate of a ROC-AUC score was returned to the teams, calculated on a small subset of files. The final AUC scores on the whole test set were given after the deadline of the contest. Throughout the paper, we will report AUC scores, and also global accuracy values (ACC), true positive and negative rates (TPR, TNR). 

\section{Data partition}

Three datasets recorded \textit{in situ} in different places and with different acoustic conditions were made available to the challenge participants: two for development --- Freefield1010 (FF) and Warblr (W) --- and one for the final evaluation. These datasets are comprised of 10-second 16-bit 44.1 kHz audio recordings that were manually labeled with binary labels indicating the presence/absence of bird sounds at file-level. Manual labels were available for FF and W, with about 25\% of positive samples (files with bird sounds) for FF and 75\% in the case of WW. For more details about these data, the reader may refer to the challenge Website\footnote{\url{http://machine-listening.eecs.qmul.ac.uk/bird-audio-detection-challenge/}} and the challenge description article~\cite{stowell2016bird}. 

Table \ref{tab:data} shows how the two development datasets were randomly split into three subsets, in number of files. The Train/Valid/Test proportions were respectively 0.8/0.05/0.15 for FF and  0.85/0.0625/0.0875 for W. Preliminary experiments were carried out on each dataset separately and on the merge subsets FF\_W\_1. Then, the final models were trained on FF\_W\_2, in which the train and test subsets were merged. 


\section{Features}

As input to the networks, 56 log-Mel triangular-shaped filter-bank (F-BANK) coefficients were extracted every 50 ms on 100 ms duration frames, with 50 Hz and 22050 Hz as minimum and maximum extreme frequency values to compute the Mel bands, respectively. Hence, for each 10 second file, a $200\times 56$ matrix was extracted. 
This matrix is used as a single input image fed to the networks. 

Other types of features were tested: delta and delta-delta added to the static F-BANK coefficients, 13 or 56 Mel-Frequency Cepstral Coefficients, raw FFT coefficients (dimension: $430\times 512$), fingerprints (dimension: $192\times 200$), etc. All these features were extracted with different resolutions, but in the end the set of 56 static F-BANK features always performed better.

Several pre-processing techniques were also tested, such as global mean removal, mean and variance standardization, ZCA whitening, but no gain and even sometimes performance decreases were observed compared to raw F-BANK features. In particular, centering the data based on the average spectrogram computed over the training set did not help.

\section{Models \label{sec:models}}

Several types of CNNs were tested: standard CNNs, residual CNNs (resNets), and densely connected CNNs (denseNets). If we denote the convolution operation of the $i^{th}$ layer by a function $F_i$ of the output of the preceding layer $x_{i-1}$, the output $x_{i}$ of the $i^{th}$ layer is given by the following equations for the three model types:

\begin{itemize}
\item standard CNNs~\cite{lecun1995convolutional}: $ x_{i}=F_i(x_{i-1}) $
\item resNets~\cite{he2016deep}: $x_i=F_i(x_{i-1})+x_{i-1}$
\item denseNets~\cite{huang2016densely}: $x_i=F_i([x_0, x_1, \ldots, x_{i-1}])$
\end{itemize}

In the literature, performance was shown to degrade when adding too many layers in standard CNNs due to the increased difficulty to train very deep CNNs~\cite{he2016deep}. ResNets were created to build successful very deep models by modeling the residuals of the input instead of the input itself. DenseNets are an extension of resNets, in which the outputs of the preceding convolution layers are concatenated rather than summed to the current layer output, within so-called "dense blocks"~\cite{huang2016densely}. To ensure shape consistency between the layers used in residual or dense blocks, a stride of one and the convolution mode 'same' were used. One advantage of denseNets lies in the fact that they achieve comparable or even better performance than resNets with orders of magnitude less parameters. Indeed, in preliminary experiments, denseNets with 328K parameters outperformed standard CNNs comprised of 1-2M parameters and performed as well as resNets of 4M weights. For this reason, denseNets were the preferred solution for the challenge.


Figure \ref{fig:densenet} shows the architecture details of the denseNets used in this work. They are comprised of an initial convolution layer (Conv) followed by three dense blocks of five batch-normalization/ReLu/Conv layers each, each block followed by a transition block comprised of $1\times 1$-Conv-max-Pooling layers. In the dense blocks, squared $3\times 3$ convolution and $2 \times 2$ pooling filters were used. Decisions are taken after a global mean-pooling layer and a dense output layer with a Softmax non-linearity activation function. The total number of layers and parameters are 74 and 328K, respectively.  Theano~\cite{2016arXiv160502688short} and Lasagne were used to perform the experiments on two GPUs: a TITAN X and a 1080 devices.

\begin{figure*}
\centering
\includegraphics[scale=0.68]{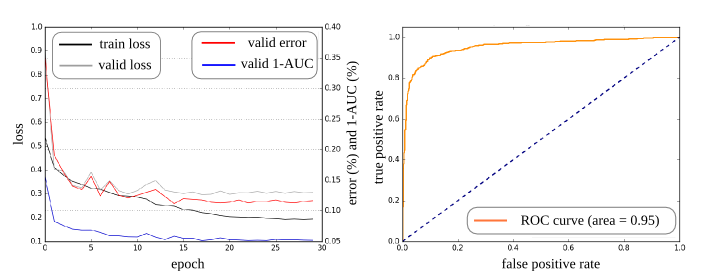}
\caption{Left: training and valid losses (left y-axis), valid error and 1-AUC scores (right y-axis) as a function of the training epochs. Right: ROC curve on the 884-sample valid subset. The two figures were obtained with one of the best denseNet models.}
\label{fig:training-curves}
\end{figure*}

\section{Experiments}

\subsection{Training setup}

Preliminary experiments were carried out on each dataset separately and then on the merged datasets as reported in Table \ref{tab:data}. For instance, it appeared that FF contains samples more difficult to classify than W. CNNs trained on FF gave an AUC score of 91.6\% on the FF Test subset, while the same kind of models trained on W reached a 95.2\% score on the W Test set.

The FF and W subsets were then merged into the FF\_W\_1 corpus as referred to in Table \ref{tab:data}, in order to optimize several hyperparameters: the number of convolution layers, residual and dense blocks, the filter sizes and number of feature maps, the learning rate (LR) and its decay strategy, the mini-batch size (10 samples), the number of training epochs (30 epochs), the data augmentation methods (time and frequency randoms shifts, described in the next Section), etc. In particular, LR was tuned through random search on a log-scale~\cite{bergstra2012random}. Parameter updates were performed with Nesterov momentum updates with a standard 0.9 momentum value optimizing a categorical cross-entropy cost function. A binary cross-entropy objective, although theoretically equivalent to its categorical counterpart used in multiclass problems, led to slightly worse performance, and so did a binary hinge cost function as well. Hence, we kept two outputs in all the models. The best learning rate strategy has been to divide by a factor of 2 the learning rate after the first eight training epochs with a fixed LR of 0.019326. 

Figure \ref{fig:training-curves} illustrates training monitoring curves (loss on the train and valid subsets, error rate and (1-AUC score) on the valid subset as a function of the 30 total training epochs) and the ROC curve calculated on the valid subset. A plateau is reached after about 10 epochs as shown by the loss and error curves on the valid subset. On the contrary, the loss curve on the train set continues to decrease after 10 epochs showing that slight overfitting still occur. A 95\% AUC ROC score was obtained on the valid subset, as shown in the right-hand side of the figure.
 
Finally, the FF\_W\_1 train and test subsets were merged into FF\_W\_2, totaling 14,806 files used to train the best models. 


\subsection{Data augmentation}

\begin{table}
\centering
\caption{Comparison between data augmentation strategies} 
\label{tab:dataaugment}
\begin{tabular}{lcccc}
 & AUC(\%) & ACC(\%) & TNR(\%) & TPR(\%) \\
\midrule
no-augment & 95.2$\pm$0.1  & 89.9$\pm$0.2 & 92.9$\pm$0.2 & 86.0$\pm$0.3 \\
cropping & 95.7$\pm$0.1 & 90.7$\pm$0.2 & 93.5$\pm$0.1 & 87.1$\pm$0.3 \\
time-reversing & 94.2$\pm$0.1 & 88.3$\pm$0.2 & 96.0$\pm$0.2 & 78.4$\pm$0.3 \\
crop. + time-rev. & 95.3$\pm$0.1 & 89.4$\pm$0.2 & 91.8$\pm$0.2 & 86.4$\pm$0.3 \\
\bottomrule
\end{tabular}
\end{table}

Data augmentation refers to artificially enlarging the training dataset using label-preserving transformations in order to reduce overfitting. In image processing tasks, a very simple and common form of data augmentation consists in generating image reflections and image translations using zero-padding and cropping, both transformations performed on-the-fly during training only, which has the advantage of avoiding to store the transformed images on disk~\cite{krizhevsky2012imagenet}. In audio tasks, cropping pixels in the spectrogram is equivalent to time and frequency shifts, and a reflection or horizontal flip corresponds to reversing the time axis. Both transformations were investigated in this work.

In this work, cropping consisted in the following: 1) the input images are zero-padded with 2 pixels at both extremities of the time and frequency axes, 2) two integers to be used as offset values in both dimensions are randomly drawn from a uniform distribution with possible output values of 0, 1 and 2. Corresponding time and frequency shift values are 0 ms, 50 ms or 100 ms, and 0 Hz, 66.7 Hz or 133.4 Hz, respectively. It should be noted that no tests were made to optimize the upper limit of 2 pixel shifts. Reflection (time-reversing), when activated, was randomly applied on-the-fly during training with a 0.5 probability drawn for each new incoming training sample.

Results on the FF\_W\_1 test subset are given in Table \ref{tab:dataaugment}: without data augmentation, with cropping only, with time-reversing only and with both cropping and time-reversing. A first model was trained on the FF\_W\_1 train subset with no data augmentation and the randomly initialized initial model was saved to be used in the next data augmentation conditions. This ensures model initialization has no impact on the result comparison between the data augmentation strategies. There is still some randomness due to the random shuffle of the training samples at each epoch. The 95\% confidence intervals given in the table were estimated nonparametrically by bootstrap sampling with a hundred samples of size 1000 data points.

As shown in the table, cropping only is best, with 95.7\% AUC and 90.7\% ACC scores compared to 95.2\% AUC and 89.9\% ACC obtained without data augmentation. Interestingly, time-reversing degrades performance by 1.0\% and 1.6\% absolute in both AUC and ACC scores, compared to not using data augmentation. Time-reversing seems to significantly help in the detection of the negative class (no-birds, TNR: 96.0\%, +3.1\%) but degrades even more the detection of the positive class (birds, TPR: 78.4\%, -7.6\%). This may be explained by the fact that the temporal structure of bird singing is meaningful for the model and randomly reverting time loses this structure. 

\subsection{Experiments on the official test set}

\begin{figure}
\centering
\includegraphics[scale=0.36]{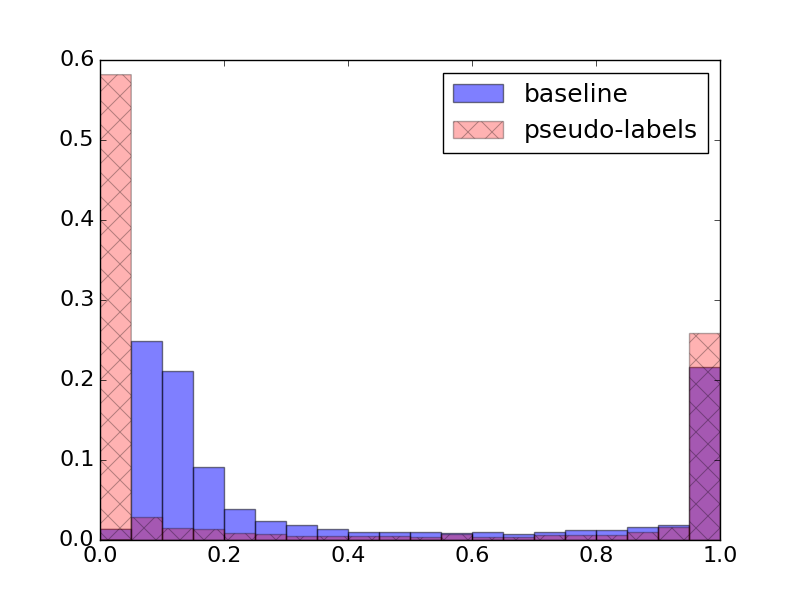}
\caption{Probability normalized histograms for a denseNet trained on FF\_W\_2 (baseline) and one trained with FF\_W\_2 enlarged with test predictions (pseudo-labels).}
\label{fig:hist}
\end{figure}

In this section, comparisons between interesting submissions are explained, and the strategies that led to the final score of 88.22\% (rank: $3^{\textrm{rd}}/30$ participants) are described. It is interesting to observe that performance consistently decreased of about 0.5\% when comparing the preview and the final scores. For instance, the best submission scored 88.79\% and 88.22\% on the preview and final test sets, respectively. This drop in performance is a sign of lack of generalization power of the models together with a possible effect of the small size of the preview test subset in number of samples. In the remainder, only the final scores are reported.

All the submissions were based on denseNets, except one based on a CNN. The CNN submission led to the worst AUC score of 85.46\%, far from the 87.41\% score obtained with a baseline DenseNet. Using as much training data as possible proved to be useful: training a denseNet on FF only gave a 86.29\% to be compared to a 87.41\% score, when using FF\_W\_2 for training. 

The following ideas led to improvements over the denseNet 87.41\% baseline:

\begin{itemize}
\item Fine-tuning on adversarial training samples,
\item Model parameter averaging over the highest scoring training epochs, 
\item Ensemble probability averaging.
\end{itemize}

A non-negligible proportion of training files were mis-classified by baseline denseNets: about 1200 files corresponding to 8\% of the 14.8K training files. This was due to the fact that some files had been incorrectly labeled by the human annotators (manual annotation was crowdsourced), but also to the fact that some files were more difficult to classify than others, due to noisy conditions, predominance of insect  or water sounds, etc. Thus, one idea has been to fine-tune models on these difficult training samples. Using the same learning rate as the one used with the whole training dataset proved to be too harsh but a learning rate 3-order magnitude smaller than the original one led to small improvements. Only 10 fine-tuning training epochs on the 1,200 files were performed and the accuracy on these files increased from 0\% to about 40\%. 
Although statistically not significant, a slight gain of about 0.1\% absolute in the final test scores was obtained with this fine-tuning strategy. 

During training, the usual way to select the final model parameters is done by saving the model that achieved the best performance on a valid subset. This could be called the "single-best-epoch" models. Another common idea consists of averaging the model parameters over several training epochs.  This is known to be usually helpful as it can be seen as a kind of parameter regularization. These models are expected to benefit from a larger generalization power. In our case, we observed that even if the global score on the Valid subset increases during training, there are fluctuations in the true positive and true negative rates between two successive epochs, so that parameter averaging was expected to be beneficial. Indeed, parameter averaging over the best epochs brought about 0.2-0.3\% absolute gains compared to using "single-best-epoch" models. A minimum AUC threshold of 94\% on the valid subset was used to select the best training epochs. In general, about 20 epochs out of the 30 training epochs usually verified this criterion.  

Finally, the third successful idea has been to perform ensemble averaging on the output probabilities provided by several models. The final best submission with a 88.22\% AUC score was obtained by averaging the prediction probabilities of four different models, using the geometric mean. These were obtained by using two models with different weight random initializations, with parameter averaging over the best epochs, together with their two counterparts fine-tuned on the mis-classified files of the training set. The idea to take the geometric mean came from preliminary experiments, in which it consistently outperformed other averaging methods such as arithmetic and harmonic means. This is an interesting result that needs to be further studied.


Several ideas revealed unsuccessful but one is maybe more worth to be reported: the idea of using predictions on the official test set as pseudo-groundtruth labels to enlarge the training dataset with data from the test set. Only the most confident predictions were kept: the ones with probabilities below 0.3 or greater than 0.7. With these thresholds, about 7800 files out of 8620 test files were selected and merged to the 14.8K training files. New denseNets were trained on this augmented dataset but a significant decrease in the AUC score was observed, with values around 87.2\%. What was notable with these models has been the change in the probability distributions on the test set that were much less spread than the ones of the previous models, as shown in Figure \ref{fig:hist}. In this figure, two normalized histograms are represented: the one with red bars filled with an 'x' pattern corresponds to the probabilities outputted by a model trained on the enlarged training data, and the one with plain blue bars to a denseNet trained on FF\_W\_2 only. It appears that the first histogram is much more peaky around the two extreme probabilities 0 and 1. Pseudo-labeling had the effect to push the two probability modes, usually around 0.2-0.3 and 0.7-0.9, towards the extrema 0 and 1, which seems satisfying in terms of confidence measures but unfortunately led to more errors in the end. More selective probability thresholds such as 0.1 and 0.9, and also majority voting, were tested, but these led to the same worse results. 




\section{Saliency maps \label{sec:saliency}}

Recently, several methods have been proposed in the literature to compute so-called "saliency maps" with a trained neural network~\cite{zeiler2014visualizing,simonyan2013deep,springenberg2014striving}. Saliency maps are a visualization of which input pixels are important for the model to make a prediction. The common idea of these approaches consists of computing the gradient of the network prediction for a given input sample with respect to the input. The best approach was reported to be the one by Springenberg called \textit{guided backpropagation}~\cite{springenberg2014striving}. 

\begin{figure}
\centering
\includegraphics[width=\linewidth]{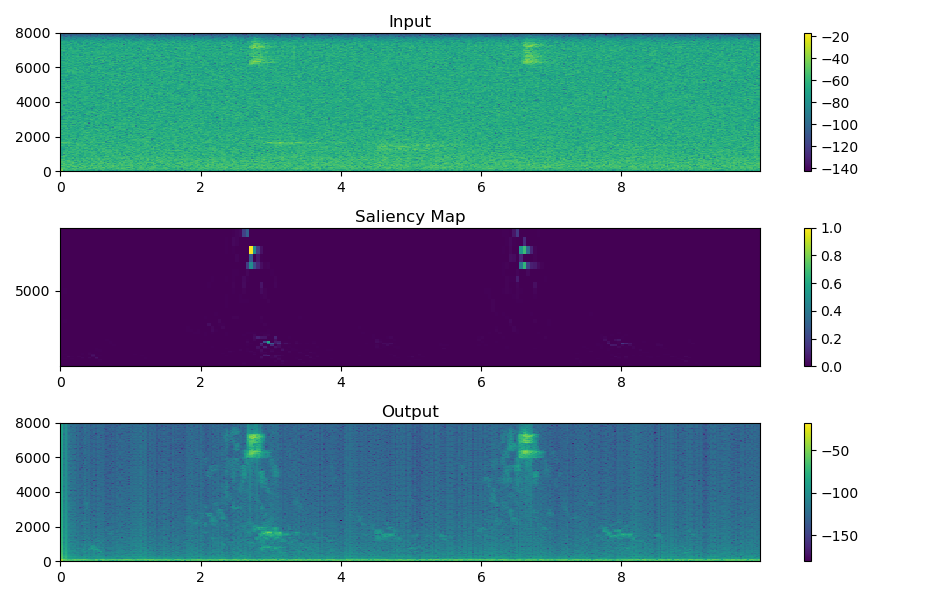}
\caption{Example of a noisy file of the positive class. Top: original spectrogram, middle: saliency map, bottom: saliency-masked output spectrogram.}
\label{fig:saliency}
\end{figure}

With this method, the saliency maps of the test samples were computed to illustrate which parts of the F-BANK coefficient time series were important to predict if birds sounds were present.  Figure \ref{fig:saliency} is an example on a file of the positive class. Two small salient blobs clearly appear on the saliency map, and they indeed correspond to bird sound components.

Besides simple illustration purposes, a saliency map can be multiplied to the Fourier spectrogram of the original audio sample to enhance the crucial frequency bins supposed to belong to bird sounds. Since 56 F-BANK coefficients were used as input features to the models, we adapted the saliency maps by simply assigning the saliency value of a given F-BANK coefficient to all the frequency bins contained in the frequency range of the filter bank coefficient. Then, the saliency-masked spectrogram is inversed back to the time domain by an overlap-add synthesis through inverse FFT. Audio samples can be listened to online\footnote{Audio samples available at \url{https://www.irit.fr/~Thomas.Pellegrini/}}.

\section{Conclusions}

In this paper, experiments using several types of convolutional neural networks (i.e. standard CNNs, residual nets and densely connected nets) were reported in the framework of the BAD challenge. DenseNets were the preferred solution since they were the best performing and most compact models, leading to a 88.22\% area under the ROC curve score on the test set of the challenge (rank: 3/30). Performance gains were obtained thank to data augmentation through time and frequency shifting, model parameter averaging during training and ensemble methods using the geometric mean. On the contrary, attempts to enlarge the training dataset with samples of the official test set with the automatic predictions used as pseudo-groundtruth labels consistently degraded performance. Further experiments in this semi-supervised setting need to be conducted, for instance one may try to use a hat-shaped loss penalty in order to penalize pseudo-labels with low confidence~\cite{zhu2005semi}.  

\bibliographystyle{IEEEtran}
\bibliography{IEEEabrv,bad2016topel}
%



\end{document}